\documentclass[12pt]{iopart}

\usepackage[dvips]{graphicx}

\begin{document}

\title[The problem of timing noise]{Searching for gravitational
waves from the Crab pulsar - the problem of timing noise}
\author{Matthew Pitkin and Graham Woan}
\address{Department of Physics and Astronomy, Kelvin Building,
University of Glasgow, Glasgow G12 8QQ, UK}
\ead{matthew@astro.gla.ac.uk}

\begin{abstract}
Of the current known pulsars, the Crab pulsar (B0531+21) is one of
the most promising sources of gravitational waves. The relatively
large timing noise of the Crab causes its phase evolution
to depart from a simple spin-down model. 
This effect needs to be
taken in to account when performing time domain searches for the
Crab pulsar in order to avoid severely degrading the search efficiency.
The Jodrell Bank Crab pulsar ephemeris
is examined to see if it
can be used for tracking the phase evolution of any gravitational wave signal
from the pulsar, and we present a method of heterodyning the data that takes account
of the phase wander. The possibility of
obtaining physical information about the pulsar from comparisons
of the electromagnetically and a gravitationally observed timing
noise is discussed. Finally, additional problems caused by pulsar
glitches are discussed.

\end{abstract}

\maketitle

\section{Introduction}
Of the known isolated pulsars the Crab pulsar (B0531+21) is
thought to be one of the most promising sources of gravitational
waves. This is due to its relative youth ($\sim{1000}$ years) and
 correspondingly large spin-down rate. The criterion
which makes the Crab pulsar such a promising source also provides
one of the major problems faced for search algorithms designed to
detect it, that of timing noise. Timing noise has been known about
since the early days of pulsar observations, and represents a
random walk in phase or frequency of the pulsar about the regular
spin-down model (Cordes and Helfand, 1980). It is thought to be intrinsic to
the pulsar and not just a result of magnetospheric perturbations
or interstellar propagation. Several different models for its origin have been suggested
(Cordes and Greenstein, 1981) although its origin is still unknown. Timing noise has been shown to be
well correlated with spin-down rate (Arzoumanian \etal, 1994), and therefore it is no surprise
that
the Crab pulsar has a large timing noise
component. Many other pulsars are seen to have timing noise, but
it is negligible for almost all of those within the frequency band of the LIGO and
GEO600 interferometers.

\section{The Crab pulsar ephemeris}
The Crab pulsar has been extensively studied since 1969 and its
behaviour from 15 February 1982 to the present is recorded in a
monthly ephemeris produced at Jodrell Bank
({\it http://www.jb.man.ac.uk/pulsars/crab.html}). This
contains the pulse arrival time at the solar system barycentre
(given as the arrival time of the peak of the first pulse after
midnight on the given day), the frequency ($\nu$) and first frequency
derivative ($\dot{\nu}$) of the pulsar. The phase residuals caused by timing noise can be seen in the
Jodrell Bank ephemeris after removing the third order Taylor expansion of the phase,
\begin{equation}\label{spindown}
\phi = \phi_0 + \nu_0(t-t_0) + \frac{1}{2}\dot{\nu_0}(t-t_0)^2 +
\frac{1}{6}\ddot{\nu_0}(t-t_0)^3,
\end{equation}
from that calculated using the ephemeris
data (see figure 1).
There is a good agreement between figure 1 and
phase residual plots 
in figure 4 of Lyne \etal (1993) and Scott \etal (2003)
indicating that the ephemeris is sufficient to track the Crab
pulsar phase. Any difference between figure 1 and those in the above references is most
probably due to different lengths of data used for the fitting procedure. Lyne \etal (1993) showed that
the timing noise does not vary significantly on scales of less
than a month, meaning that the ephemeris should have sufficient
time resolution.

\begin{figure}
\begin{center}
\includegraphics[width=0.75\textwidth]{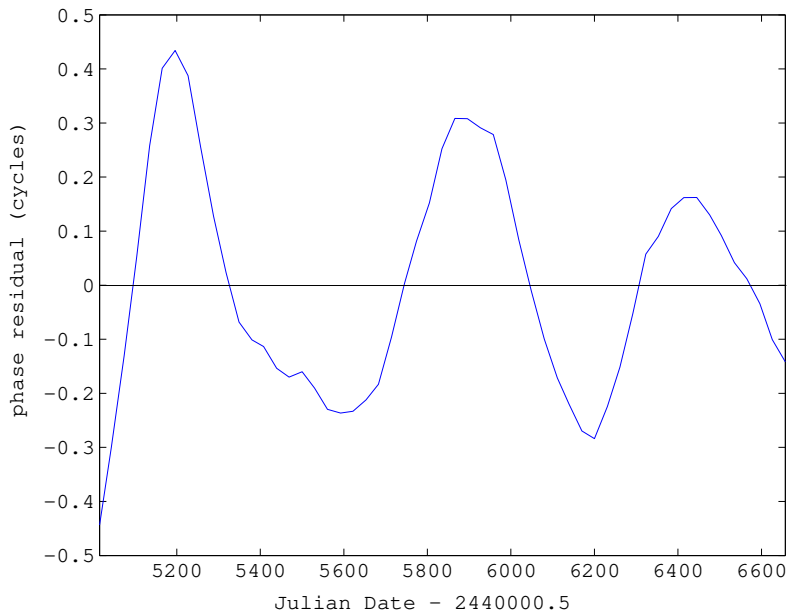}
\end{center}
\caption{\label{timingnoise}The effect of timing noise as seen in the phase residual of the Crab pulsar
obtained by removing a third order fit from the phase as
calculated using the Jodrell Bank Crab pulsar ephemeris.}
\end{figure}

\section{Search method}
Current search methods for detecting gravitational waves from
known pulsars, as developed for the LIGO and GEO600 analyses (The LIGO Scientific
Collaboration 2003),
require an accurate knowledge of their phase evolution, 
and any timing noise has to be correctly accounted for. 
The gravitational wave signal from a pulsar as seen at the
detector (Jaranowski \etal, 1998) is defined as
\begin{equation}
h(t) = F_+ (t,\psi) h_0 \frac{1+\cos^2 \iota}{2} \cos \Phi(t) +
F_\times (t, \psi) h_0 \cos \iota \sin \Phi(t),
\end{equation}
where $F_+$ and $F_\times$ are the detector's beam pattern
dependent on the gravitational wave polarisation angle $\psi$,
$\iota$ is the angle between the pulsar's spin direction and the
propagation direction of the gravitational waves and $\Phi(t)$ is the
phase evolution of the pulsar including Doppler effects from the
Earth's rotation and orbital motion. The time domain search method developed by Dupuis 
and Woan (in preparation) and used in LIGO Scientific Collaboration (2003) uses
heterodyning procedures to remove the time varying parts of $\Phi$
(as determined through radio observations). This leaves a
gravitational wave signal that would vary only with the detector
beam patterns.
The analysis method uses Bayesian inference to infer
posterior probability densities for the unknown pulsar
parameters of $h_0$, $\iota$, $\psi$ and $\phi_0$ (the pulsar's
phase at reference time $t_0$).
This method is fine for pulsars with small timing noise, and would
be fine for the Crab pulsar over short integration periods of
days, but as signals from pulsars will be weak very long
integration times of months to years are needed. The timing noise 
in the Crab pulsar will cause the phase to drift
significantly from a simple spin-down model over a relatively short time,
thus degrading the efficiency of or completely nullifying the
search method. Therefore the timing noise has to be
dealt with by applying an extra heterodyne that removes its
effects. Here the ephemeris data needs to be used to track the
phase. The complete phase evolution can be calculated by applying
a spline fit between consecutive monthly frequencies in the
ephemeris using the frequency, its first derivative and the phase
as boundary conditions. This produces a fifth order equation for
the phase of the form
\begin{equation}\label{spindown2}
\phi = \phi_0 + \nu_0(t-t_0) + \frac{1}{2}\dot{\nu_0}(t-t_0)^2 +
\frac{1}{6}\ddot{\nu_0}(t-t_0)^3 +
\frac{1}{24}\ddot{\nu_0}\dot{}(t-t_0)^4 + \frac{1}{120}\ddot{\nu_0}\ddot{}(t-t_0)^5,
\end{equation}
in which the spin-down parameters $\ddot{\nu_0}$, $\ddot{\nu_0}\dot{}$ and
$\ddot{\nu_0}\ddot{}$ vary on a monthly basis.
The phase as calculated in equation (\ref{spindown2}) is then removed in the extra heterodyning
stage of the analysis,
\begin{equation}
\textrm{heterodyned data} = \textrm{data} \times \exp(-\phi_{\textrm{timing noise}}).
\end{equation}
 It is acknowledged that this interpolation method is maybe not the most accurate fitting
technique although it seems adequate for our purposes. 

\section{Discussion and conclusions}
Timing noise may not just be a nuisance that gets in the way of
our search method, but may provide some astrophysical
insight into the structure of neutron stars. It is generally
assumed that the neutron star and its magnetosphere are strongly
coupled. Any difference in the timing noise phases
between gravitational and electromagnetic radiation will provide
information on the strength of this coupling. To test this
coupling a new parameter, $\alpha$, can be added into the search
algorithm of Dupuis and Woan. This parameter has been investigated by Jones
(private communication) and is defined as the scaling factor between
the electromagnetic and gravitational phase noise, reflecting changes in
the moment of inertia and magnetosphere of the neutron star. This
information would be useful in constraining the various equation
of state models of neutron stars.

As well as timing noise the Crab pulsar, like many other pulsars, is also
subject to glitches. These are characterised by a sudden increase
in the pulsar frequency and spin-down rate, followed by a slower
decay back to close to their pre-glitch values. These glitches also need to
be taken into account in searches, in order to avoid the uncertain phase
evolution around them decreasing the search efficiency. 
It is unknown what effect glitches
could have on the gravitational wave phase, due to possible changes in the moment of
inertia or crustal structure of the pulsar, so that after each
glitch a new initial phase parameter will be needed in the search
algorithm to take into account a possible jump in phase. This is equivalent to
performing coherent searches only between glitches, and then combining the results
incoherently.

The Crab pulsar is not unique in its large timing noise or glitchiness, and as
gravitational wave
detectors develop better sensitivity at lower wavelengths the
methods highlighted above will need to be further developed and
applied more often in pulsar analyses. This in turn will require close
observation in radio wavelengths to characterise the phase
evolution. In cases where radio observations are not regularly available
comparisons should be made of different methods of fitting to the data that
provide the best extrapolation results. It is also worthwhile to study the
effects of timing noise and how it effects the search efficiency
and coherence times
when thinking about searches for unknown pulsars, the most
promising candidates for which are young and therefore many are
likely to have large timing noise. We are requesting additional Crab
timing data from Jodrell Bank on a shorter timescale with which to test the
spline interpolation and give a more precise tracking of the phase.

The prospect of getting physical information about pulsars from
the comparisons of electromagnetic and gravitational timing noise makes these considerations
particularly important.

\section*{References}
\begin{harvard}
\item[]Arzoumanian Z \etal 1994 {\it Ap. J.} {\bf 422} 671-680
\item[]Cordes J M and Helfand D J 1980 {\it Ap. J.} {\bf 239}
640-650 
\item[]Cordes J M and Greenstein G 1981 {\it Ap. J.} {\bf
245} 1060-1079 
\item[]Dupuis R and Woan G {\it in preparation}
\item[]Jaranowski P \etal 1998 {\it Phys. Rev. D} {\bf 58}
63001
\item[]Jones D I {\it private communication} 
\item[]Jodrell Bank Crab Pulsar Timing: Monthly Ephemeris {\it http://www.jb.man.ac.uk/research/pulsar/crab.html} 
\item[]Lyne A \etal 1993 {\it Mon. Not. R. Astron. Soc.} {\bf 265} 1003-1012
\item[]Scott D M \etal 2003 {\it Mon. Not. R. Astron. Soc.} {\bf 344} 412-430
\item[]The LIGO Scientific Collaboration; Abbott B \etal 2003
Setting upper limits on the strength of periodic gravitational
waves using the first science data from the GEO600 and LIGO
detectors {\it Preprint} gr-qc/0308050 

\endrefs

\end{document}